%% file: thesismod.tex
\documentclass[twoside]{ruthesis}
\include{epsf}
\newcommand{\gsi}{\,\raisebox{-0.13cm}{$\stackrel{\textstyle>}
{\textstyle\sim}$}\,}
\newcommand{\lsi}{\,\raisebox{-0.13cm}{$\stackrel{\textstyle<}
{\textstyle\sim}$}\,}

\begin{document}
% Use \draft if until you are done
%\draft

%  Title and other sections that come before the body  of the document

\include{putthesisfront}
    \include{titlepage}
    \include{thesis-titlemod}
    \afterpreface %  This commmand must always be present before body
		  %  of text.
% Now lets include the body of the document...

\include{thesis-chap1mod}

\include{thesis-chap2mod}

\include{thesis-chap3mod}

\include{thesis-chap4mod}

\include{thesis-chap5mod}

% In the appendix some things should change, so a user should say that
% its starting.
    \appendix
    \include{thesis-appendixmod}

%    \bibliography{thesis}
%    \bibliographystyle{unsrt}

%    \include{thesis-vita} 
\end{document}

%% file: putthesisfront.tex
\rightline{hep-ph/9612337}
\rightline{December, 1996}
\baselineskip=18pt
\vskip 0.6in
\begin{center}
{ \LARGE The Electroweak Phase Transition in the Minimal Supersymmetric Standard
Model}\\
\vspace*{0.6in}
{\large Marta Losada\footnote{On leave of absence
from Centro Internacional
de F\'{\i}sica and Universidad Antonio Nari\~{n}o, Santa Fe de
Bogot\'a, COLOMBIA.}} \\
\vspace{.1in}
{\it Department of Physics and Astronomy \\ Rutgers University,
Piscataway, NJ 08855, USA}\\
\vspace{.2in}
\vspace{.1in}
\end{center}
\vspace*{0.05in}
\vskip  0.2in  
%\eject
%\vspace*{.8in}

Abstract:  Using dimensional reduction we construct an effective 3D theory of the
 Minimal Supersymmetric
Standard Model at finite temperature. The final effective theory is obtained
after three successive stages of integration out of massive particles. We obtain the
full 1-loop relation between the couplings of the reduced theory and the underlying
4D couplings and masses. The
procedure is also applied to a general 
two Higgs doublet model and the Next to Minimal Supersymmetric Standard Model.
 We analyze the effective 3D theory  constructed for the MSSM to determine
the regions of parameter space for which electroweak baryogenesis is possible. We find
that the inclusion of all supersymmetric scalars has the effect of enhancing the strength
of the phase transition. The requirement of a very light stop is not necessary for 
baryogenesis. The phase transition is sufficiently first order if
the lightest Higgs mass, $m_{h}\lsi 70$ GeV. We note the existence of potentially
interesting regions of parameter space for which existing analysis techniques are
inadequate to decide the question.
 
\thispagestyle{empty}
\newpage
\addtocounter{page}{-1}
%\tableofcontents
%\listoffigures
\newpage

%% file: thesis-titlemod.tex
%  Title and other sections that come before the body  of the document

% For a  PhD give the command \phd. If you don't do this then the default
% is for a Masters Degree.
% Optional fields are (where normal is for PhD or the default Master
% of Science:
%	\degree (normally Doctor of Philosophy or Master of Science)
%	\draft (this should probably be set  until near the final
%              drafts. Currently it modifies the title page.)
    \phd
 %   \jointumdnj
%   \draft
    \title{\textbf{The Electroweak Phase Transition in the Minimal Supersymmetric
Standard Model}}
    \author{ Martha Losada}
    \campus{New Brunswick}
    \program{Physics and Astronomy}
    \director{Glennys R. Farrar}
    \approvals{4}
    %\copyrightpage % Do you want copyright protection?
    \submissionmonth{October}
    \submissionyear{1996}
    \figurespage
    %\tablespage
    \abstract{\input{thesis-abstractmod}}  % required for Ph.D.
\beforepreface 
% The \beforepreface  command actually causes insertion of the title, 
% abstract,  and copyright pages into the new document.

   % \preface{%
%Most of chapters 1 through 6 of this document are quoted almost verbatim from
%the Manuscript Form \cite{styleguide} published by the Graduate School--New 
%Brunswick, however,  some changes have been made. This is primarily
%due to the assumption (implicit in the Graduate School document) that
%a typewriter or simple word processor is being used to prepare the
%document. Because  the form of this sample dissertation has been
%accepted by the Graduate School any of the inconsistencies between the
%two should be acceptable.
%
%The appendix first discusses how to organize the \LaTeX{} manuscript files for
%your thesis or dissertation. It also attempts to describe enough of
%\LaTeX\ that it could be used, at least initially, without the manual.
%}
    \acknowledgements{
I would like to especially thank  Glennys Farrar  for proposing this line
of research, as well as many inspiring discussions  on the physics related to
this thesis and comments on the 
manuscript.
 
\noindent
I  am also indebted to Mikhail Shaposhnikov whom I thank for many useful discussions.

%\noindent
%I especially appreciate the support I have received from my family. In particular,
%my husband and my parents. 

\noindent
Research supported in part by Colciencias, Colombia.
}
  %  \dedication{A Rafa}
   % \abbreviations{\input{thesis-abbreviations}}
% The \afterpreface  command actually causes insertion of the
% contents, list of figures, etc. into the new document.

%% file: thesis-abstractmod.tex
 Using dimensional reduction we construct an effective 3D theory of the
 Minimal Supersymmetric
Standard Model at finite temperature. The final effective theory is obtained
after three successive stages of integration out of massive particles. We obtain the
full 1-loop relation between the couplings of the reduced theory and the underlying
4D couplings and masses. The
procedure is also applied to a general 
two Higgs doublet model and the Next to Minimal Supersymmetric Standard Model.
 We analyze the effective 3D theory  constructed for the MSSM to determine
the regions of parameter space for which electroweak baryogenesis is possible. We find
that the inclusion of all supersymmetric scalars has the effect of enhancing the strength
of the phase transition. The requirement of a very light stop is not necessary for 
baryogenesis. The phase transition is sufficiently first order if
the lightest Higgs mass, $m_{h}\lsi 70$ GeV. We note the existence of potentially
interesting regions of parameter space for which existing analysis techniques are
inadequate to decide the question.

%% file: thesis-chap1mod.tex
\chapter{Introduction}
\hspace*{2em} The main goal of the research presented in this thesis is to analyze the electroweak phase transition in the context of
the Minimal Supersymmetric Standard Model (MSSM). The principal motivation resides in the implications of the
phase transition on the study of the baryon asymmetry of the universe.
It has been shown, that unless the phase transition is sufficiently first order, electroweak scale
physics cannot account for the observed asymmetry \cite{kuzmin}.

In chapter two we present an overview of baryon number violation and its relation to the
electroweak phase transition. In chapter three we introduce the formulation of finite
temperature field theory. The contents of these two chapters are well known. The purpose is to point out
to the reader the main issues related to the baryon asymmetry which inspire the study of the phase transition.

 The Standard
Model has been fully investigated in relation to these issues. The analysis of this model 
depends on only
one unknown parameter, the Higgs mass. 
A complete study of the phase transition must
address the problem posed by the infrared divergences which are a characteristic of gauge theories at finite
temperature. In order to do this non-perturbative methods must be implemented. 
The most accurate calculations have ruled out the Standard Model
as the responsible for the baryon asymmetry for any value of the Higgs mass \cite{KLRS}.

Extensions of the Standard Model may provide desirable features that affect the conclusions
about the electroweak phase transition. In particular, the MSSM contains additional
particles which could significantly change certain aspects of the analysis of the strength
of the phase transition. The number of
unknown parameters is increased, and a detailed study of their effects must be performed.

In chapter 4 we construct an effective theory of the MSSM requiring  the presence of 
a single light Higgs at
the phase transition. This allows us to use the constraints, obtained by lattice calculations,
 on the first
order phase transition
for the effective 3D theory of the Standard Model. The fifth chapter explores the MSSM parameter space to determine
the regions for which the phase transition is sufficiently first order.

Chapters 4 and 5, as well as the results presented in Appendices B, C and F are
original contributions to this problem. The contents of chapter 4 and appendices A-D have appeared
in reference \cite{mlosada1}.

%% file: thesis-chap2mod.tex
\chapter{Overview of Baryon Number Violation and the Electroweak Phase Transition}

%% file: thesis-chap3mod.tex
\chapter{Finite Temperature Field Theory}

%% file: thesis-chap4mod.tex
\chapter{High Temperature Dimensional Reduction}

\hspace*{2em} See hep-ph/9605266.

%% file: thesis-chap5mod.tex
\chapter{The Electroweak Phase Transition in the MSSM}
\section{Introduction}

\hspace*{2em}  Many different authors have studied the order of the
electroweak phase transition in the Minimal Supersymmetric Standard Model (MSSM). 
Most of these studies relied on a one- and two-loop finite temperature effective potential 
analysis of the phase transition \cite{giudice, zwirner1, zwirner2, carena1, espinosa1} in which the stops were
expected to make the most significant contribution from supersymmetric particles. 
The authors of these studies, in the limit of a large pseudoscalar Higgs mass, $m_{A}\rightarrow \infty$, have identified a region of parameter space
for which the transition is strong enough.  This corresponds to low values of  $\tan\beta$, and values of 
the soft supersymmetry breaking right stop mass, $m_{U_{3}}^{2}$, which are small or even negative \footnote{In reference 
\cite{zwirner2} the analysis was extended for the full range of allowed values of $m_{A}$. It was  found
that larger values of $m_{A}$ are favored.}.

A different approach consists of separating
 the perturbative and non-perturbative aspects of the phase transition. This is performed
 through the perturbative construction 
of effective three dimensional theories, and a subsequent lattice analysis of the reduced theory \cite{kks, farakos, KLRS, kajantie}. 
In addition, this  also provides a check to perturbative results for the phase transition.
 One may construct effective 3D theories for different models at
finite temperature to study the electroweak phase transition. 
As shown in chapter 4, for the case in which the reduced theory
 contains a single
light Higgs field, characterized by a Higgs self-coupling, $\bar{\lambda}_{3}$, and an effective  3D gauge 
coupling, $g_{3}$, the condition for a sufficiently strong first order phase 
transition  becomes \cite{KLRS}

\begin{equation}
x_{c} = {\bar{\lambda}_{3}\over g_{3}^{2}} \lsi 0.04.
\label{xcrit}
\end{equation}
The quantity $x_{c}$ is a function of the different parameters appearing in the original 4D model.
  
The analysis of parameter space for the reduced theory of the Standard Model was  
performed in \cite{KLRS, kajantie}. The conclusion was that for no value of the Higgs mass is electroweak 
baryogenesis possible. Although inconsistent with current experimental constraints,
 purely perturbative studies had allowed electroweak baryogenesis for small enough  values of the Higgs mass.

 In chapter 4 we constructed a 3D theory for the MSSM including Standard Model particles and
 additional
corrections arising from gauginos, higgsinos and all squarks and sleptons. Additionally, one-loop
corrections from dimensional reduction to all couplings in the model were calculated. 
We use these results to  explore the MSSM parameter space
 in order to determine the regions for which electroweak baryogenesis may occur.
An important element in this analysis  is to establish the relation between
the running parameters in the original 4D theory and physical parameters.  This is done in
appendix E. 
In section 5.2 we discuss our numerical results from scanning parameter space. In section 5.3 we conclude.

In references \cite{cline, laine} 3D theories
for the MSSM have recently been constructed, and  analyzed to determine the regions of
parameter space where the criteria given by equation (\ref{xcrit}) is fulfilled. In  these
papers only  the contribution from gauge bosons, higgses and third generation quarks and squarks to the
 3D reduction was included. Nor did they incorporate one-loop corrections
to all of the parameters in the theory.
In our work all one-loop corrections have been included, as well as contributions from all
SUSY particles. This allows us to investigate the effect of  extra supersymmetric particles, 
in addition to third generation squarks, on
the strength of 
the phase transition.
Furthermore the results of \cite{cline, laine} are not totally in agreement. In reference \cite{laine} the results
agreed basically with those found in the perturbative effective potential analysis.
The most favorable region of parameter space was found to be $m_{h}\lsi m_{W}$ (low $\tan\beta$), small stop mixing, 
$m_{U_{3}}\lsi 50$ GeV and $m_{A} \gsi 200$ GeV.
In addition to this region, reference \cite{cline}  found another region of parameter space in which 
 arbitrary values of $\tan\beta$
and a range of values for the pseudoscalar Higgs mass,  $40\lsi m_{A} \lsi 80$ GeV, give a sufficiently
strong phase transition.

\section{Discussion of Numerical Results}

\hspace*{2em} As mentioned above, the 
quantity  $x_{c}$ becomes a function of
the  parameters in the model: $x_{c} = x_{c}(M_{A}, m_{o}, \mu, m_{{1\over 2}}, m_{\tilde{g}},
 A, \tan\beta,T_{c})$, as well as of the gauge couplings.
  $A$  is the trilinear soft SUSY breaking
parameter, taken to be universal. $\mu$ is the supersymmetric mass parameter. $ m_{o}, m_{{1\over2}}, m_{\tilde{g}}$  denote the
common squark/slepton mass at the SUSY breaking scale, the  $SU(2)$ gaugino  and  gluino mass  respectively.
We take $M_{A}$ to be the physical pole mass of the pseudoscalar Higgs, and $\tan\beta $ is the 
ratio of the vacuum expectation values of the Higgs fields in the renormalized zero temperature
theory. The ratio in
$x_{c}$ also depends indirectly on the scale $M_{SUSY}$. $M_{SUSY}$  is the scale at which
we have assumed a universal mass parameter for squarks and sleptons, as well as the scale at which
the SUSY boundary conditions on the quartic Higgs couplings  appearing in the Higgs potential
are imposed \cite{hempfling}. 
 All of the above mentioned quantities are our input parameters chosen in such a way that experimental and 
theoretical constraints are satisfied. In addition, in order to study the effect of the masses
of third generation
squarks we keep the stop soft supersymmetry breaking masses $m_{Q_{3}}, m_{U_{3}}$, as independent parameters. 

We define the critical temperature 
$T_{c}$, from the requirement of the existence of a direction in field
space
at the origin of the Higgs potential for which the transition to the minimum of the potential
in the broken phase  
can occur classically.
In the 3D lattice calculations \cite{farakos, KLRS} the critical temperature is defined
by the temperature at which phase coexistence disappears. In general, these two values of temperature
are close. In fact, the actual value of the critical temperature
lies between these two values. We will remark later on the circumstances under which there can be a significant
difference arising from this distinction.

The procedure  to evaluate $x_{c}$ is the following:\\
- restrict initial parameter space by experimental constraints on the masses of the particles,\\
- calculate the critical temperature for a fixed set of parameters, \\
- check the validity of the high temperature expansion, \\
- check the adequate suppression of non-renormalizable terms and the
validity of perturbation theory,\\
- determine the value of $x_{c}$.

Throughout our analysis  we will concentrate on the regions of parameter space which describe an effective
 theory 
in which there is a single light scalar and thus the bound given by equation (\ref{xcrit}) is valid. However, we 
mention that another possibility is the scenario in which two scalars, e.g. one Higgs and additionally a right stop,
are both nearly massless at $T_{c}$ \cite{carena1}.

\subsection{Dependence on $\tan\beta$ and $M_{A}$}

\hspace*{2em} As is well known and is shown in appendix E, we can parametrize the Higgs sector in terms of two quantities: 
$\tan\beta$ and the pole mass of the pseudoscalar Higgs boson, $M_{A}$.
The range of variation explored for these input parameters is taken  as follows:\\
- we allow $M_{A}$ to vary between 40-300 GeV. The lower bound is taken from
experimental considerations, the upper limit to insist on the validity of
the high-temperature expansion.\\
- we vary $\tan\beta$ between $1.25$ and $13.3$. For values of $\tan\beta$ outside of this
range the results have qualitatively  the same behaviour \footnote{In the dimensional reduction
procedure, the explicit dependence on 
all Yukawa couplings was kept. For our numerical
analysis,
except for the top Yukawa coupling,
we will take the value of these couplings to be zero. It might be thought that for large values of
$\tan\beta$ the bottom Yukawa coupling can also be relevant. We have explicitly checked that
this is not the case.}.

In general, the masses of all particles are taken such that the high temperature expansion
is valid.
The experimental constraints we impose on the masses are: for stop masses
 $m_{\tilde{t}_{2}} \gsi 50$ GeV,  $m_{\tilde{t}_{1}} \gsi m_{t}$,
 for first and second generation squarks $m_{\tilde{q}_{i}} \gsi 200$ GeV,
 sleptons $m_{l} \gsi 50$ GeV, the gluino
mass either $\lsi 1$ GeV or  $\gsi 150$ GeV \cite{abe, farrar}. In addition, the value of the
left soft supersymmetry breaking stop mass $m_{Q_{3}}$ must be such that the contribution from stops
and sbottoms to the $\rho$ parameter is not too large \cite{zwirner1}.

The critical temperature $T_{c}$, is evaluated from the temperature dependent Higgs mass matrix as explained in
chapter 4. The requirement of a zero eigenvalue of this mass matrix will define
the direction in field space for which the curvature of the potential at the origin is zero.
We shall make a few general remarks of the dependence of the value of the critical temperature with
respect to the input parameters in the model. As the value of $\tan\beta$ decreases,
 the critical temperature 
also  decreases. The dependence on the pseudoscalar Higgs mass is not very strong, but as $M_{A}$ increases 
the critical temperature decreases. For a given
value of $\tan\beta$, the value of $T_{c}$ varies at most on the order of 5 GeV  as  $M_{A}$ takes on values in the range mentioned above.
 The only other parameters which significantly affect the critical
temperature are the masses of the squarks/sleptons.  With 
respect to the dependence of critical temperature on the squark masses, we note in
particular that
 a lower value of the right stop supersymmetric breaking mass increases the value of the critical
 temperature. 
We mention that we have checked that the difference in the critical temperature from the diagonalization of
 the Higgs mass matrix,
 equation (10) in \cite{mlosada1} and from equation (7.9) in \cite{laine} is extremely
 small ($\leq .1$ GeV) and for our purposes
negligible.

We have placed all of the plots of the results of the analysis of the strength of the phase
transition into appendix F.
Figure \ref{xctanbeta} shows the value of $x_{c}$, for the case of no squark mixing,
 as a function of the pseudoscalar Higgs mass for  values
of $\tan \beta$ ranging from 1.25 to 13.3. We have fixed the other parameters to be
$m_{o}=50$ GeV, $m_{{1\over 2}}= 50$ GeV,
 $m_{\tilde{g}}= {\alpha_{s}\over \alpha_{W}}m_{{1\over 2}}$, $M_{weak}= m_{t}$,
$M_{SUSY} = 10^{12}$ GeV \footnote{Note that in our approximation the values of the first and 
second generation squarks and the slepton masses are fixed by $m_{o}$, $m_{\tilde{g}}$, $m_{{1\over 2}}$,
through the renormalization group running, and
are constant as we vary $\tan\beta$ and $M_{A}$. The left and right stop masses change as we move
on the curves plotted in figure \ref{xctanbeta}.}.
For large values of $\tan \beta$, there is no value of $M_{A}$ for which $x_{c}$  fulfills 
the condition given by equation (\ref{xcrit}), and $x_{c}$ varies very little as you vary
$M_{A}$. However, for low values of  $\tan \beta$ and  large enough values of $M_{A}$, $x_{c}$  can be small enough for the phase
transition to be sufficiently first order. 
The strong dependence of $x_{c}$ on the value of the pseudoscalar Higgs mass, for 
low values of the ratio of the vacuum expectation values, arises basically through the dependence
of the quantity $\bar{\lambda}_{3}$ in equation (\ref{xcrit})  on the mixing angle,
 see equation (17) in \cite{mlosada1}. It is easy
to see to lowest order the same dependence on $M_{A}$ arising in  finite temperature effective potential analysis \cite{zwirner2}.

\subsection{Dependence on other parameters}

\hspace*{2em} We now discuss the consequences of the variation of the other parameters in the model on the strength of the phase transition.
Figures \ref{xcMsusy} show $x_{c}$ vs. $M_{A}$, for  $\tan\beta = 1.25,1.75,13.3$,
and for two different values of the SUSY scale. 
The value of the SUSY scale makes an important change in $x_{c}$.
 This is expected because the weak scale values  for the masses and couplings  
change when we change the SUSY scale (see appendix E). In particular, the  value
of the quartic Higgs couplings, which largely
determine the value of $\bar{\lambda}_{3}$ in equation (\ref{xcrit}),
is changed very much as we vary the SUSY scale.

We plot in figure \ref{xcmo} the dependence of $x_{c}$  on
 the soft-supersymmetry breaking mass $m_{o}$, for  $\tan \beta =1.25, 1.75$. We show the results for 
two values of the soft supersymmetry breaking mass for the squarks/sleptons, the solid line corresponds
to $m_{o}=50$ GeV, the dotted line to $m_{o} = 150$ GeV. 
 We see that, keeping the gluino
and $SU(2)$ gaugino mass fixed, the strength of the phase transition
 becomes weaker for a larger value of $m_{o}$. However, we cannot lower the value of $m_{o}$ much due to experimental constraints on the 
squark/slepton masses.
The dependence of the critical temperature and of $x_{c}$ on  the mass of the $SU(2)$ gaugino
is very small. When we increase the value of the gluino mass  by an amount of  order of $50$ GeV, 
 it produces a small variation in  the value of the critical temperature,  and  changes the value of
$x_{c}$ by an extremely small amount.

We plot in figure \ref{xcmu}  the variations of $x_{c}$
for different values of $\mu$  and the $A$ parameter.
The solid line corresponds to no mixing, the dashed line to $\mu =200$ GeV, and the dotted
line to $A= 150$ GeV. A small value of $\mu$ is favored in most of parameter space while changing
$A$ makes a negligible effect.

In figure  \ref{xcmt} we plot the influence of the mass of the top, which implies a change in
the top Yukawa coupling, on the strength of  the phase transition. We show the effect
for two values of $\tan\beta=1.5,1.75$, where the dashed line corresponds
to $m_{t}=165$ GeV, the solid line to $m_{t}=175$ GeV, and the dotted line to $m_{t}=190$ GeV.
 The same dependence on top mass was
observed in \cite{laine}.
The results of varying the right stop soft supersymmetry breaking
mass, $m_{U_{3}}$, are shown in figure \ref{xcmoRU}. We plot the values of
$x_{c}$ as a function of the pseudoscalar Higgs mass for $\tan\beta = 1.25,1.5,1.75$, and $m_{U_{3}} =
50,100,200$ GeV.  We see that we recover the expected 
dependence on $m_{U_{3}}$ \cite{zwirner2, carena1, laine}: lowering $m_{U_{3}}$ causes
 $x_{c}$ to decrease.

We  have also compared the results of our general analysis, to simplifying cases which did not include
the effects from all supersymmetric particles. In all cases we have kept the full one-loop
corrections to the 3D couplings.
In figures \ref{comp1} and \ref{comp2} we plot $x_{c}$ vs. $M_{A}$ for $\tan\beta= 1.25,1.5.1.75,13.3$.
The solid line corresponds to the general case including all supersymmetric particle contributions
to the dimensionally reduced theory. The dotted line represents the case in which gluino and $SU(2)$ gaugino
contributions are neglected. In figure \ref{comp1}, we allow
 the running stop masses to vary as  functions of $\tan\beta$ and $M_{A}$. In figure \ref{comp2}
the values of the running stop masses are fixed.
Figure \ref{comp3},
 shows the variation of $x_{c}$ with $M_{A}$ for three
different cases. The solid line corresponds to our general analysis, as given above. The dashed line is
for the case in which only the effect of third generation squarks is included. The gluino/gaugino thermal
screening contribution to the 3D masses of the squarks
is also excluded. The dotted line corresponds to the case in which we  include  all of
other squarks and sleptons, ignoring all
gluino and gaugino contributions to the three dimensional theory. In order to compare the approximations,
the masses of the squarks
and sleptons have been fixed to be the same in all three cases. As expected, the dependence on the values of 
$\tan\beta$ and $M_{A}$ is very similar in each case. We can see that as a result of including the contributions of all scalars the strength of the
first order phase transition is enhanced. We have checked for all cases that the dependence on the value of
the right stop soft supersymmetric breaking mass has the same effect of decreasing the value of 
$x_{c}$.   However, we see that for the case in which the effect
of all squarks and sleptons is included the change induced in the strength of the phase transition
is not so large as for the case in which only the contribution from third generation
squarks \cite{laine} was  considered. 
We have also compared the two cases in which only third generation squarks were included with and without
thermal screening arising from the gluino and gaugino. The differences in the values of $x_{c}$ for this case
are negligible.

\subsection{Validity of approximations}

\hspace*{2em} Previous treatments of this problem \cite{zwirner1, zwirner2, espinosa1, laine} were content
with the result of the dependence of the strength of the phase transition on $\tan\beta$ and
$M_{A}$.
We wish to emphasize however that for some regions of parameter space it may not be correct
to conclude from this analysis that the phase transition is not sufficiently strongly first order.
In particular, for some regions of parameter space $x_{c}$ is an extremely strong function
of temperature near the critical temperature. This occurs when
 the  mixing angle $\theta$, which diagonalizes the 3D Higgs mass matrix at finite temperature,
varies rapidly in  certain regions of temperature. In fact,  what is happening is that the diagonal elements of the 3D Higgs mass matrix are
becoming equal for a certain value of the temperature. Nevertheless, the two eigenvalues of the mass matrix
differ substantially close to $T_{c}$. That is, one the eigenvalues is much larger than the other, the latter 
becoming equal to zero at the critical temperature. This indicates that our procedure for integrating
out the heavy Higgs doublet is correct.
If the critical temperature for the phase transition is 
close to the value of the temperature where this rapid variation occurs,
 then the value of $\theta$ and consequently of $x_{c}$
 will vary exceedingly close to $T_{c}$.

The value of the temperature at which this rapid variation occurs depends strongly
on the value of the pseudoscalar Higgs mass. As $M_{A}$ increases this
 temperature also increases. In addition, for larger values of
$\tan\beta$ the mixing angle dependence on the temperature is less strong than for
low values of $\tan\beta$. We show in figures \ref{thetaT40}, \ref{thetaT300} the dependence of
$\theta$ on the temperature. In figure  \ref{thetaT40}  we have fixed the value of $M_{A}=40$ GeV,
 where the solid line corresponds to  $\tan\beta= 1.25$
and the dashed line to $\tan\beta= 13.3$, keeping all other parameters fixed. Figure \ref{thetaT300}
is the same thing for $M_{A}=300$ GeV. 
In figure \ref{xcT40}, we plot the value of $x_{c}$ as a function of temperature close
 to $T_{c}$ for $M_{A}=40$ GeV.
We see that a variation on the order of $5$ GeV in the temperature induces a change in the value
 of $x_{c}$, $\Delta x_{c}\sim .13$.
For the case in which $M_{A}= 300$ GeV, the variation around $T_{c}$ implies a variation of $\Delta x_{c} \sim 0.005$.
We conclude that possibility of  large uncertainty in the mixing angle is relevant only for low values of
$\tan\beta$ and  $M_{A}$.

Low values of $\tan\beta$ imply a larger value of the top Yukawa coupling. This in turn, 
decreases the value
of $m_{U_{3}}^{2}$, which could even take on negative values.
 However, we cannot analyze this situation as it
 implies a breakdown of our approximations. A very light right stop
cannot be integrated out in the construction of the effective theory.
 The effect of the gluino mass is to increase the values of
the running soft supersymmetry breaking mass parameters for squarks as the energy scale decreases.
For a given  value of $m_{o}$, the masses of the running squarks and sleptons will
be much smaller as the gluino mass decreases.
Consequently, 
a light gluino case, in which the masses of the gluino and $SU(2)$ gaugino are taken to be zero, 
can easily accomodate the scenario originally proposed in \cite{carena1} in which there 
are a light Higgs and a light right stop
at the phase transition. 

\section{Conclusions}

\hspace*{2em} To summarize, we  conclude that  a sufficiently strong electroweak phase transition
 can be fulfilled in the 
MSSM for values of $\tan\beta \lsi 1.75$. The value of $x_{c}$  decreases as the pseudoscalar mass
increases.
The pseudoscalar Higgs mass can be as low as $M_{A}=100$ GeV, depending on the value of $\tan\beta$,
and still give the required $x_{c}\leq 0.04$.  For 
low values of $\tan\beta$ and $M_{A}$ the rapid variation of the mixing angle of the 3D Higgs mass matrix,
as a function of the temperature, does not allow us to conclude definitely whether it is possible
to have a sufficiently strong phase transition. A precise determination of the critical temperature
is needed in this case which requires consideration of tunneling.
The inclusion of all supersymmetric scalars has the effect on enhancing the strength of the
phase transition. In particular, it is not necessary to have a
very light right stop for equation (\ref{xcrit}) to be fulfilled. That is, in contrast to purely
pertubative analysis, having
universal soft supersymmetry breaking masses at the SUSY breaking does not impede the phase transition
from being sufficiently first order.
 The effect of the parameters $A$, $\mu$, $m_{\tilde{g}}$, $m_{{1\over 2}}$,
$m_{t}$ is  either small or increases the value of $x_{c}$. Using the results presented in appendix E
we can obtain the values of the lightest Higgs boson mass.
For the allowed region of parameter space, we observe that the corresponding upper bound on
 the lightest physical
 Higgs mass, $m_{h} \lsi 70$ GeV, is very close to being experimentally accesible.

%% file: thesis-appendixmod.tex
\chapter{ MSSM in Four Dimensions}

\hspace*{2em} See hep-ph/9605266.

\chapter{Explicit Relationships between Parameters}

\hspace*{2em} See hep-ph/9605266.

\chapter{2HDM and NMSSM}

\hspace*{2em} See hep-ph/9605266.

\chapter{Appendix of Finite Temperature Formulae}

\hspace*{2em} See hep-ph/9605266.

\chapter{Relation to Physical Parameters}

\hspace*{2em} The one-loop relations of the
running parameters in the $\overline{MS}$-scheme to the parameters of the effective 3D theory are given
in appendix B.
Additionally, the running parameters in the $\overline{MS}$-scheme should be given with the
same accuracy in terms of physical parameters. This requires one-loop renormalization
of the 4D zero temperature theory.
In the construction of the effective 3D theory we started out with a 
4D Lagrangian with running parameters at the scale $\mu_{4}$. Generically, 
we can express  any coupling or mass parameter of the dimensionally reduced 3D theory
in terms of the running parameters by

\begin{equation}
F_{3D} = f(\mu_{4}) - \beta_{f}^{b} L_{b} -  \beta_{f}^{f} L_{f},
\label{TRG}
\end{equation}
(plus additional temperature squared dependent terms for the three dimensional mass terms),
 where $L_{b}$ and $L_{f}$ are defined in appendix D.
 The superscripts
$b$, $f$ denote the bosonic and fermionic contributions respectively, to the one-loop beta function of the
corresponding parameter.
All scale dependence, implicit
in the quantities $L_{b}$ and $L_{f}$, of the 3D 
parameters will drop out when we relate the running parameters to physical
variables.
 
For the masses and couplings we use the zero temperature one-loop renormalization group equations
to relate the values of the parameters at the scale $\mu_{4}$ to their values at the weak scale.
We approximate  the beta function coefficients  to be scale independent. Ignoring 
particle decoupling we can then write,

\begin{equation}
f(M_{weak}) = f(\mu_{4}) - (\beta_{f}^{b} +  \beta_{f}^{f}) \log({\mu_{4}\over M_{weak}}).
\label{4dRG}
\end{equation}
Substituting equation (\ref{4dRG}) into equation (\ref{TRG})
 the explicit dependence on $\mu_{4}$ is eliminated and the
relevant logarithmic ratio is $M_{weak}/T$.

We now discuss the  relevant couplings and masses which we are concerned with and which must
be treated separately.

- The strong gauge coupling, $g_{s}$, only enters through the expressions for the beta-function
coefficients. 
We use $\alpha_{s}(m_{t}) = 0.12$.

- The parameters $\mu$ and $A$ are not universal at one-loop. We do not try to fix them
in terms of physical parameters. However, we do require they satisfy experimental and
theoretical constraints. $A$ must not be too large in order to avoid 
colour symmetry breaking. $\mu$ is the higgsino mass in the unbroken phase and must
satisfy the  high temperature expansion criterion.

-  Even though in principle we could fix exactly the top Yukawa coupling
in terms of pole masses, since the value of the top mass is not known exactly we will not include the
finite corrections. The top Yukawa coupling is taken to be

\begin{equation}
f_{t}(M_{weak}) = m_{t}(M_{weak}) \sqrt 2/(\sin \beta v).
\label{topyuk}
\end{equation}

- We obtain the value of the quartic Higgs couplings at the weak scale
 by running the $SU(2)$ gauge coupling from
its measured value at the weak scale,  $g(M_{weak})= {2\over 3}$, up
to the SUSY breaking scale $M_{SUSY}$. The Higgs self coupling constants $\lambda_{i}$ at 
$M_{SUSY}$ are obtained imposing the boundary conditions given in \cite{hempfling}.
Finally, we obtain  the values of $\lambda_{i}(M_{weak})$ by integrating the renormalization
group equations
from the SUSY breaking scale to the weak scale. With this procedure we incorporate the
one-loop leading logarithmic corrections to the weak gauge and quartic couplings.

- We have assumed a common soft supersymmetry breaking mass term $m_{o}$, for squarks and sleptons
at the SUSY breaking scale. Using the renormalization  group  equations
 which have the form of equation (\ref{4dRG}), we obtain
$m_{Q_{i}}$, $m_{U_{i}}$, $m_{D_{i}}$ at the weak scale. Here $m_{Q_{i}}$, is the soft supersymmetry breaking 
left squark/slepton
mass, and $i$ denotes a family index. $m_{U_{i}}$ is the right handed soft supersymmetry breaking up-type squark/slepton mass and 
$m_{D_{i}}$ is the right  handed soft supersymmetry breaking down-type squark/slepton mass. 
  We then
require that the physical squark/slepton masses $m_{\tilde{q}_{i,1}}$,  $m_{\tilde{q}_{i,2}}$ 
satisfy the experimental constraints
mentioned in section 5.2.

The physical squark/slepton masses $m_{\tilde{q}_{i,1}}$,  $m_{\tilde{q}_{i,2}}$ are the eigenvalues
of the squared mass matrices

\begin{eqnarray}
M_{\tilde{q}}^{2} = \left( \begin{array}{cc}
m_{\tilde{q}_{L}}^{2} & m_{X_{q}}^{2} \\
m_{X_{q}}^{2} & m_{\tilde{q}_{R}}^{2}
\end{array} \right),
\label{squarkmassmatrix}
\end{eqnarray}
where $m_{\tilde{q}_{L}}^{2}$, $m_{\tilde{q}_{R}}^{2}$ are the mass terms for the left-handed and
right-handed squarks in the broken phase, while  $m_{X_{q}}^{2}$ denotes the left-right
squark  mixing parameter.  For the stops these mass parameters are given by

\begin{equation}
m_{\tilde{t}_{L}}^{2} = m_{Q_{3}}^{2} + f_{t}^{2} v_{2}^2 + {1\over 4}(g^{2} - {1\over 3}g^{\prime^{2}})(v_{1}^{2}
- v_{2}^{2})
\label{mtl2}
\end{equation}

\begin{equation}
m_{\tilde{t}_{R}}^{2} = m_{U_{3}}^{2} + f_{t}^{2} v_{2}^2 +  {1\over 3}g^{\prime^{2}}(v_{1}^{2}
- v_{2}^{2})
\label{mtr2}
\end{equation}

\begin{equation}
m_{X_{t}}^{2} = f_{t}(A v_{2} - \mu v_{1}).
\end{equation}

Using coupling and vacuum expectation values at the scale $M_{weak}$,
 $m_{\tilde{q}_{i,1}}$,  $m_{\tilde{q}_{i,2}}$ obtained from (\ref{squarkmassmatrix}) are
 not pole masses for the squarks and sleptons. However,
we cannot do better than this since actual values of
the squark and slepton masses are not known.

- In order to express the  mass terms in the Higgs potential in terms of physical parameters we 
follow the procedure of references \cite{ espinosa, carena, hempfling}
to obtain the renormalization group improved tree level potential. That is,
 in equation (22) of \cite{mlosada1}
we use 
the quartic Higgs couplings  at
the weak scale, obtained by integrating the renormalization group equations.
 Minimizing the potential
 %(\ref{potencial})
 gives 

\begin{equation}
m_{1}^{2} = - m_{3}^{2} \tan\beta - v^2 \cos^{2}\beta (2 \lambda_{1} + \lambda_{3} \tan^{2}\beta +
\lambda_{4} {\tan^{2}\beta\over 2})
\label{m12e}
\end{equation}

\begin{equation}
m_{2}^{2} = - m_{3}^{2} \cot\beta - v^2 \sin^{2}\beta (2 \lambda_{2} + \lambda_{3} \cot^{2}\beta +
\lambda_{4} {\cot^{2}\beta\over 2}),
\label{m22}
\end{equation}
where all quantities are evaluated at the weak scale.

The ratio of the vacuum expectation 
values of the neutral components of the Higgs doublets defines $\tan\beta = {v_{2}\over v_{1}}$. 

As is well known, the model contains
five physical Higgs bosons: a charged pair $m_{H^{\pm}}$, two neutral
$CP$-even scalars $m_{h,H}$ , and a neutral $CP$-odd scalar $m_{A}$ \cite{hempfling, kane, dawson}.
The running masses of the physical Higgs particles at the weak scale are given by \cite{hempfling},

\begin{equation}
m_{A}^{2} = - m_{3}^{2}(M_{weak})/\sin\beta \cos\beta
\label{mA2}
\end{equation}

\begin{equation}
m_{H^{\pm}}^{2} = - m_{3}^{2}(M_{weak})/\sin\beta \cos\beta - {1\over 2} \lambda_{4} v^{2}.
\label{mHpm2}
\end{equation}
The $CP$-even Higgs mass matrix elements are

\begin{equation}
M_{11}^{2} = - m_{3}^{2} \tan\beta + v^{2}\lambda_{1} \cos^{2}\beta
\label{M112}
\end{equation}

\begin{equation}
M_{12}^{2} = M_{21}^{2} = m_{3}^{2} + (\lambda_{3} + \lambda_{4})v^{2} \sin\beta\cos\beta
\label{M122}
\end{equation}

\begin{equation}
M_{22}^{2} = - m_{3}^{2} \tan\beta + v^{2}\lambda_{2} \sin^{2}\beta.
\label{M222}
\end{equation}
with corresponding eigenvalues given by

\begin{equation}
m_{h,H}^{2} = {1\over 2}\left( tr M^{2} \mp  \left((tr M^{2})^{2} - 4 \det M^{2}\right)^{1\over 2} \right),
\label{mhH2}
\end{equation}
where $tr M^{2} = M_{11}^{2} + M_{22}^{2}$ and $\det M^{2} = M_{11}^{2} M_{22}^{2} - (M_{12}^{2})^{2}$. The mixing
angle $\alpha$ is determined from

\begin{eqnarray}
\sin 2\alpha &=& {2 M_{12}^{2} \over \left((tr M^{2})^{2} - 4 \det M^{2}\right)^{1\over 2}} \nonumber \\
\cos 2\alpha &=& {M_{11}^{2} - M_{22}^{2} \over \left((tr M^{2})^{2} - 4 \det M^{2}\right)^{1\over 2}}.
\label{alphamix}
\end{eqnarray}
To correct for the fact that the effective potential
is defined at zero external momentum, the pole masses are obtained from the expression for
the running masses by \cite{espinosa,carena}

\begin{equation}
M_{\phi}^{2} = m_{\phi}^{2} + Re\Delta\Pi_{\phi}(M_{\phi}^{2})
\label{polemass}
\end{equation}
for $\phi = h,H,A, H^{\pm}$,  and the self-energy is defined by

\begin{equation}
\Delta\Pi_{\phi}(M_{\phi}^{2}) = \Pi(M_{\phi}^{2}) - \Pi(0).
\label{DELPI}
\end{equation}
We calculate the Higgs self-energies including corrections from top and stop loops.
These results can be found for example in \cite{carena}.

For the pseudoscalar Higgs we have

\begin{eqnarray}
\Delta\Pi_{A}(M_{A}^{2}) &=& - {3\over 16 \pi^2} f_{t}^{2}\cos^{2}\beta M_{A}^{2} F(m_{t}^{2}, m_{t}^{2},
M_{A}^{2}) \nonumber \\
&+& {3\over 16 \pi^2}  f_{t}^{2} (A\cos\beta +\mu\sin\beta)^{2} F(m_{\tilde{t}_{1}}^{2},
 m_{\tilde{t}_{2}}^{2}, M_{A}^{2})
\label{DELA}
\end{eqnarray}
and for the lightest scalar Higgs

\begin{eqnarray}
\Delta\Pi_{h}(M_{h}^{2}) &=& - {3\over 8 \pi^2} f_{t}^{2}\cos^{2}\alpha(-2 m_{t}^{2} + {1\over 2} M_{h}^{2}) F(m_{t}^{2}, m_{t}^{2},
M_{h}^{2}) \nonumber \\
&+& \sum_{i,j} {3\over 16 \pi^2} C_{h_{ij}}^{2} F(m_{\tilde{t}_{i}}^{2},
 m_{\tilde{t}_{j}}^{2}, M_{h}^{2})
\label{DELh}
\end{eqnarray}
where

\begin{eqnarray}
F(m_{1}^{2}, m_{2}^{2},  p^{2}) &=& \int_{0}^{1} dx \log {m_{1}^{2}(1-x) + m_{2}^{2}x - p^{2} (1-x)\over
m_{1}^{2}(1-x) + m_{2}^{2}x} \nonumber \\
&= & -1 + {1\over 2}\left( {m_{1}^{2} + m_{2}^{2} \over m_{1}^{2} - m_{2}^{2}} - \delta \right)\log {m_{2}^{2}\over
m_{1}^{2}} \nonumber \\
&+& {1\over 2} r \log\left[ {(1 + r)^{2} - \delta^{2}\over (1-r)^{2} - \delta^{2}}\right]
\label{fm1m2mp2}
\end{eqnarray}

with

\begin{eqnarray}
\delta &=& {m_{1}^{2} - m_{2}^{2}\over p^{2}}\\
r &=&  \left[ (1 + \delta)^{2} - {4 m_{1}^{2} \over p^{2}}\right] ^{1\over 2}.
\label{delr}
\end{eqnarray}
Additionally,

\begin{eqnarray}
 C_{h_{ij}} &=& {4 \sin^{2} \theta_{W} m_{Z}^{2}\over 3 v} \sin(\beta + \alpha) \left[\delta_{ij} +
{3 - 8 \sin^{2} \theta_{W}\over 4\sin^{2} \theta_{W}} r^{1i} r^{1j}\right] \nonumber \\
&-&  f_{t}^{2} v\sin\beta\cos\alpha \delta_{ij}
- {1\over \sqrt 2} f_{t}(A \cos\alpha - \mu \sin\alpha)(r^{1i} r^{2j} + r^{1j} r^{2i})
\label{chij}
\end{eqnarray}
where $r^{11} = r^{22} = \cos \tau$, and $r^{12} = - r^{21} = \sin\tau$.
The mixing angle $\tau$ diagonalizes the stop mass matrix (\ref{squarkmassmatrix}).
These relations and the expressions given in appendix B 
give us an expression for $x_{c}$ as a function of physical parameters.

\chapter{Figures}

\begin{figure}[h]
\vskip -100pt
\epsfxsize=4in
\epsfysize=6in
\epsffile{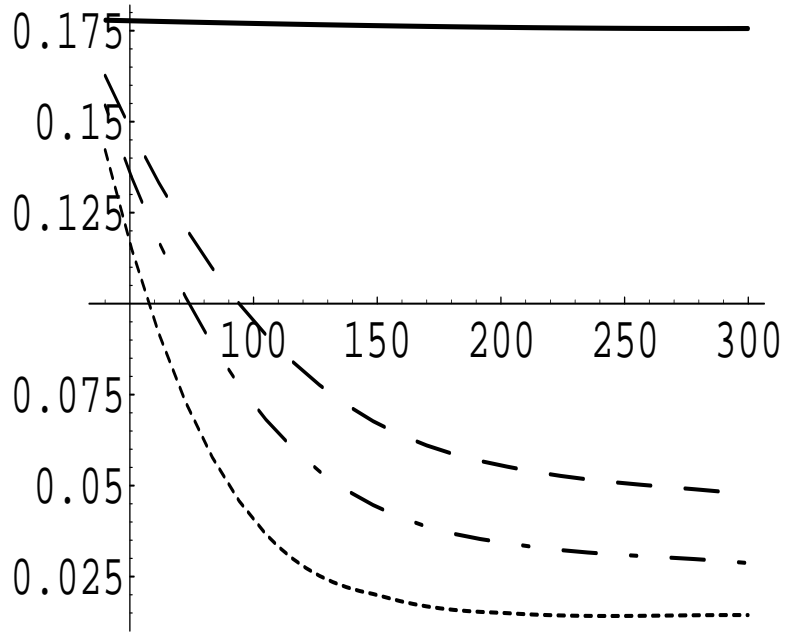}
\vskip -100pt
\caption{Plot of $x_{c}$ vs. $M_{A}$ for several different values of $\tan\beta$. The solid line
corresponds to $\tan\beta= 13.3$, the dashed line to $\tan\beta = 1.75$, the dashed-dot line to
$\tan\beta=1.5$, and the dotted line to $\tan\beta = 1.25$.}
\label{xctanbeta}
\end{figure}

\begin{figure}
\vskip -100pt
\epsfxsize=4in
\epsfysize=6in
\epsffile{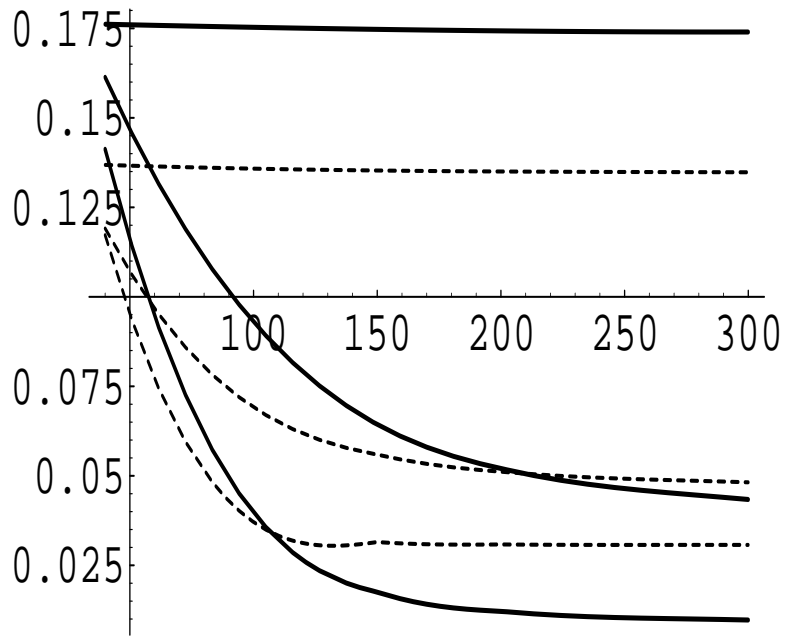}
\vskip -100pt
\caption{Plot of $x_{c}$ vs. $M_{A}$, for $\tan\beta= 1.25,1.75,13.3$. The solid line
corresponds to $M_{SUSY} = 10^{12}$ GeV, the dotted line to $M_{SUSY} = 10^3$ GeV.}
\label{xcMsusy}
\end{figure}

\begin{figure}
\vskip -100pt
\epsfxsize=4in
\epsfysize=6in
\epsffile{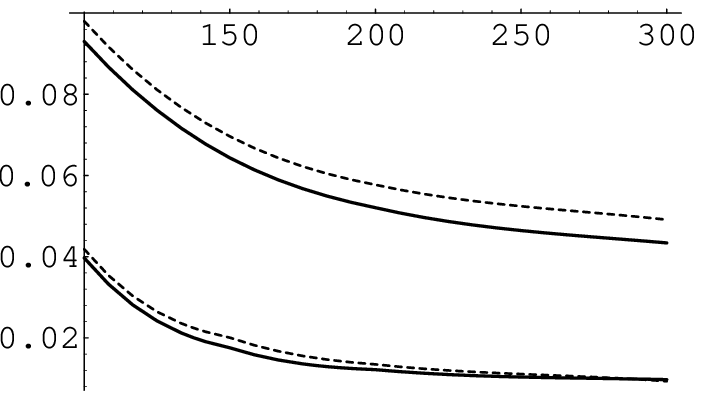}
\vskip -100pt
\caption{Plot of $x_{c}$ vs. $M_{A}$ for $\tan\beta = 1.25,1.75$. The solid line corresponds
to  $m_{o}= 50$ GeV, the dotted line to  $m_{o}= 150 $GeV.}
\label{xcmo}
\end{figure}

\begin{figure}
\vskip -100pt
\epsfxsize=4in
\epsfysize=6in
\epsffile{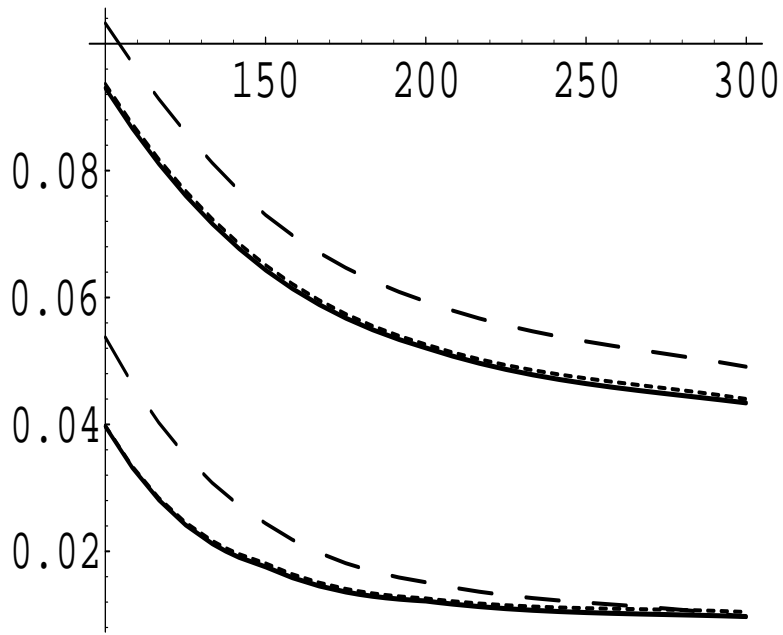}
\vskip -100pt
\caption{Plot of $x_{c}$ vs. $M_{A}$ for $\tan\beta = 1.25,1.75$. The solid line
corresponds to the case with no squark mixing, dashed line to $\mu = 200$ GeV, dotted line to $A= 150$ GeV.}
\label{xcmu}
\end{figure}

\begin{figure}
\vskip -100pt
\epsfxsize=4in
\epsfysize=6in
\epsffile{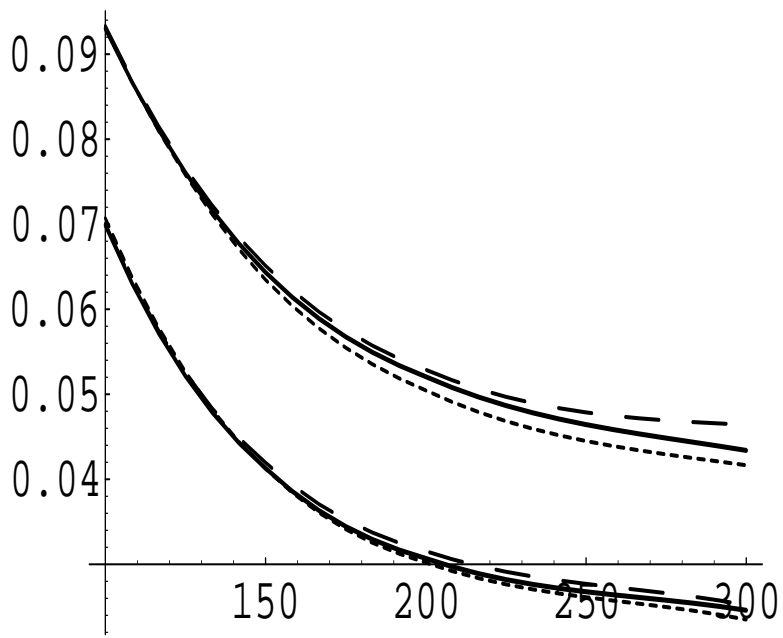}
\vskip -100pt
\caption{Plot of $x_{c}$ vs. $M_{A}$ for $\tan\beta = 1.5,1.75$. The dashed line
corresponds to $m_{t} = 165$ GeV, the solid line to $m_{t} = 175$ GeV, the dotted
line to $m_{t} = 190$ GeV.}
\label{xcmt}
\end{figure}

\begin{figure}
\vskip -100pt
\epsfxsize=4in
\epsfysize=6in
\epsffile{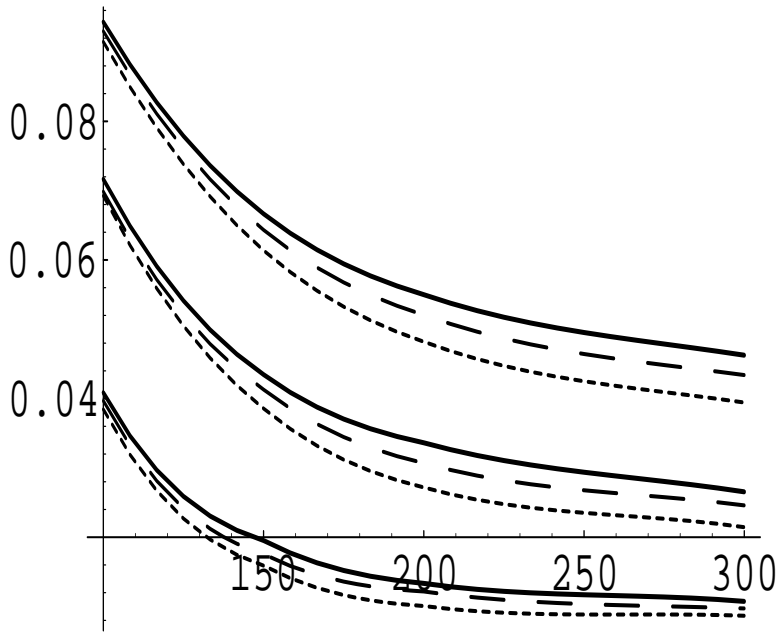}
\vskip -100pt
\caption{Plot of $x_{c}$ vs. $M_{A}$ for $\tan\beta = 1.25,1.75$. The solid line corresponds
to  $m_{U_{3}} = 200$ GeV, the dashed line to $m_{U_{3}} = 100$ GeV, the dotted line to
$m_{U_{3}} = 50$ GeV.}
\label{xcmoRU}
\end{figure}

\begin{figure}
\vskip -100pt
\epsfxsize=4in
\epsfysize=6in
\epsffile{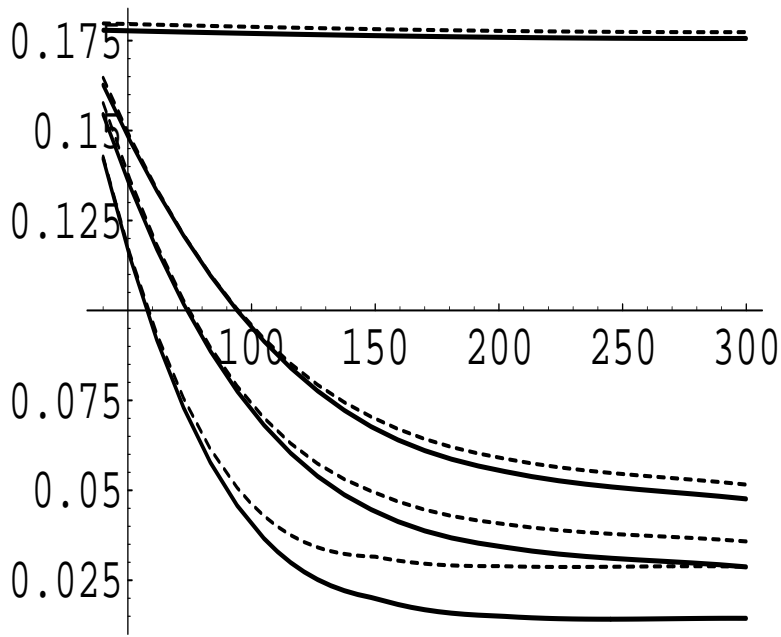}
\vskip -100pt
\caption{Plot of $x_{c}$ vs. $M_{A}$ including (neglecting) gluino contributions correspond to the 
solid (dashed) lines, for four different values
of $\tan\beta$. The stop soft supersymmetry breaking masses vary as functions of $\tan\beta$ and $M_{A}$.}
\label{comp1}
\end{figure}

\begin{figure}
\vskip -100pt
\epsfxsize=4in
\epsfysize=6in
\epsffile{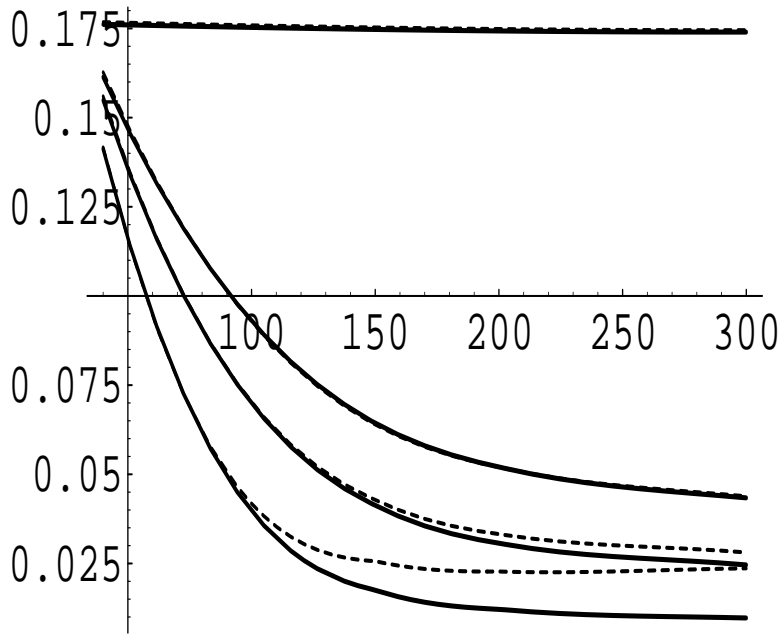}
\vskip -100pt
\caption{Plot of $x_{c}$ vs. $M_{A}$ including (neglecting) gluino contributions correspond to the
solid (dashed) lines, for four different values
of $\tan\beta$. We have fixed  the values of the stop soft supersymmetry breaking masses.}
\label{comp2}
\end{figure}

\begin{figure}
\vskip -100pt
\epsfxsize=4in
\epsfysize=6in
\epsffile{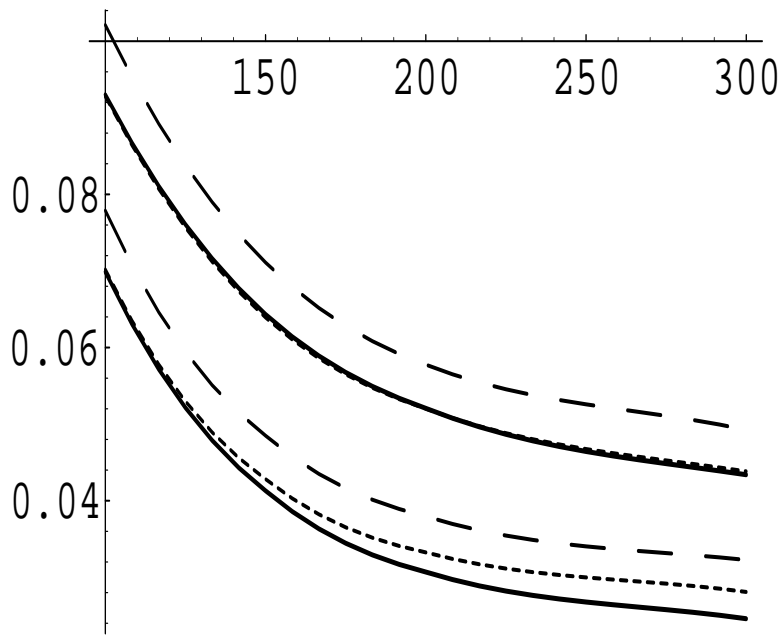}
\vskip -100pt
\caption{Plot of $x_{c}$ vs. $M_{A}$ including (neglecting) gluino contributions
 solid (dashed) lines, and
including contributions arising only from third generation squarks dotted lines,
 for two different values
of $\tan\beta$. We take fixed values of the stop soft supersymmetry breaking
 masses in all three cases.}
\label{comp3}
\end{figure}

\begin{figure}
\vskip -100pt
\epsfxsize=4in
\epsfysize=6in
\epsffile{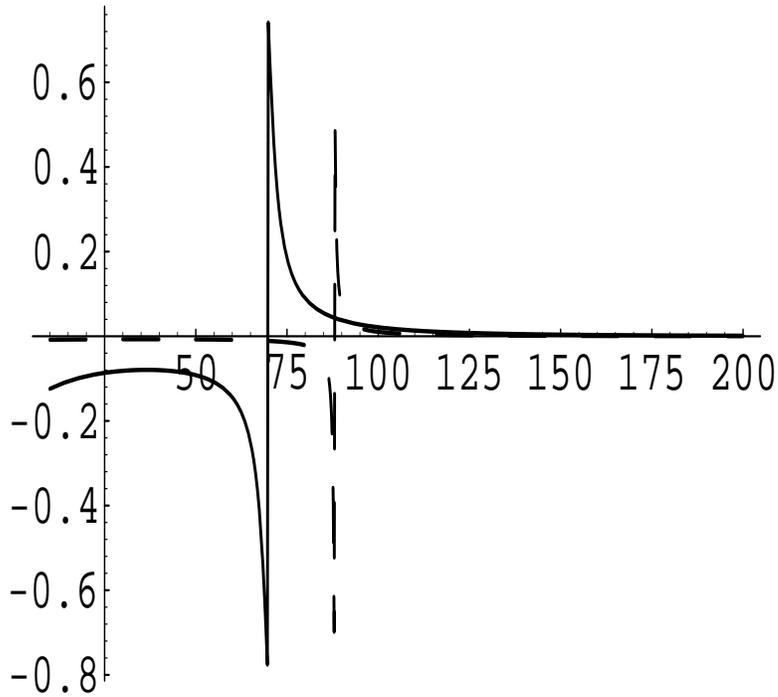}
\vskip -100pt
\caption{Plot of $\theta$ vs. $T$ for  $M_{A} = 40$ GeV. The solid line corresponds to
 $\tan\beta = 1.25$,$T_{c}= 63.3$ GeV,  the dashed line is for $\tan\beta = 13.3$,$T_{c}= 75.6$ GeV.}
\label{thetaT40}
\end{figure}

\begin{figure}
\vskip -100pt
\epsfxsize=4in
\epsfysize=6in
\epsffile{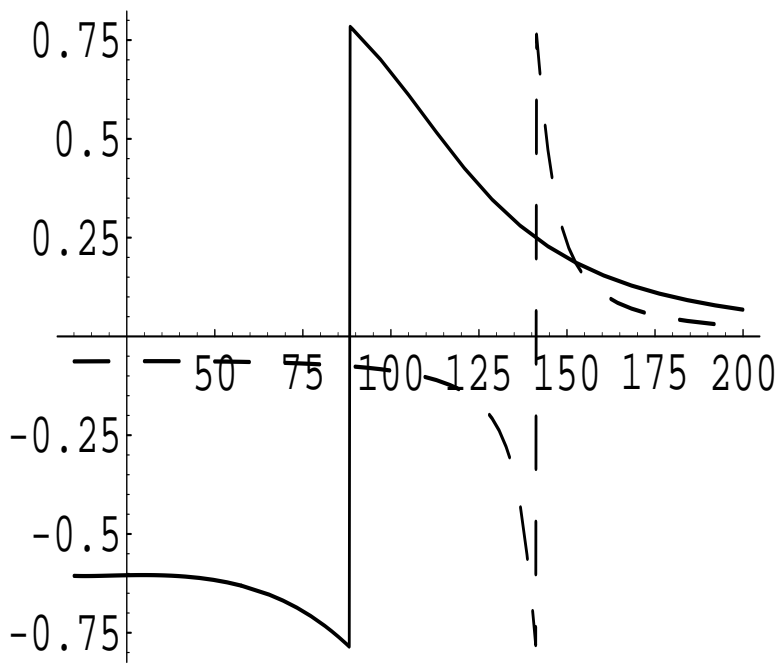}
\vskip -100pt 
\caption{Plot of $\theta$ vs.$T$ for $M_{A} = 300$ GeV, $T_{c} = 60,75.6$ GeV. The solid line
corresponds to $\tan\beta = 1.25$, the dashed line to $\tan\beta = 13.3$,  }
\label{thetaT300}
\end{figure}

\begin{figure}
\vskip -100pt
\epsfxsize=4in
\epsfysize=6in
\epsffile{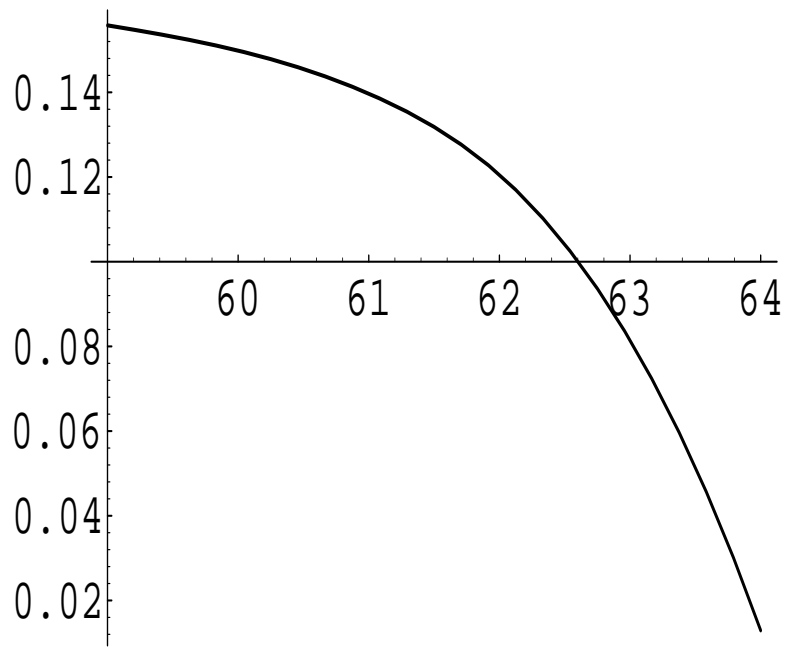}
\vskip -100pt
\caption{Plot of $x_{c}$ vs.$T$ for $\tan\beta = 1.25$, $M_{A} = 40$ GeV, $T_{c}= 60.8$ GeV. }
\label{xcT40}
\end{figure}

%% file: thesismod.bbl
\begin{thebibliography}{99}

\bibitem{kuzmin}
V.A. Kuzmin, V.A. Rubakov, and M.E. Shaposhnikov.
\newblock {\em Phys. Lett.}, B155:36, 1985.

\bibitem{KLRS}
K.~Kajantie, M.~Laine, K.~Rummukainen, and M.E. Shaposhnikov.
\newblock {\em Nucl. Phys.}, 466:189, 1996.

\bibitem{mlosada1}
M.~Losada.
\newblock Technical Report (hep-ph/9605266), 1996.

\bibitem{giudice}
G.F. Giudice.
\newblock {\em Phys. Rev.}, D45, 1992.

\bibitem{zwirner1}
J.R. Espinosa, M.~Quir\'os, and F.~Zwirner.
\newblock {\em Phys. Lett.}, B307:106, 1993.

\bibitem{zwirner2}
A.~Brignole, J.R. Espinosa, M.~Quir\'os, and F.~Zwirner.
\newblock {\em Phys. Lett.}, B324:181, 1994.

\bibitem{carena1}
M.~Carena, M.~Quiros, and C.E.M. Wagner.
\newblock {\em Phys. Lett.}, B380:81, 1996.

\bibitem{espinosa1}
J.R. Espinosa.
\newblock {\em Nucl. Phys.}, 475:273, 1996.

\bibitem{kks}
K.~Kajantie, K.~Rummukainen, and M.E. Shaposhnikov.
\newblock {\em Nucl. Phys.}, B407:356, 1993.

\bibitem{farakos}
K.~Farakos, K.~Kajantie, K.~Rummukainen, and M.E. Shaposhnikov.
\newblock {\em Nucl. Phys.}, B425:67, 1994.

\bibitem{kajantie}
K.~Kajantie, M.~Laine, K.~Rummukainen, and M.E. Shaposhnikov.
\newblock {\em Nucl. Phys.}, 458:90, 1996.

\bibitem{cline}
J.~Cline and K.~Kainulainen.
\newblock Technical Report (hep-ph/9605235), 1996.

\bibitem{laine}
M.~Laine.
\newblock Technical Report (hep-ph/9605283), 1996.

\bibitem{hempfling}
H.~Haber and R.~Hempfling.
\newblock {\em Phys. Rev.}, D48:4280, 1993.

\bibitem{abe}
F.~Abe et~al.
\newblock {\em Phys. Rev. Lett.}, 69:3439, 1992.

\bibitem{farrar}
G.R. Farrar.
\newblock {\em Phys. Rev.}, D51:3904, 1995.

\bibitem{espinosa}
J.A. Casas, J.R. Espinosa, M.~Quir\'os, and A.~Riotto.
\newblock {\em Nucl. Phys.}, B436:3, 1995.

\bibitem{carena}
M.~Carena, M.~Quiros, and C.E.M. Wagner.
\newblock Technical Report (hep-ph/9908343), 1996.

\bibitem{kane}
H.~Haber and G.~Kane.
\newblock {\em Phys. Reports}, 117:75, 1985.

\bibitem{dawson}
J.~Gunion, H.~Haber, G.~Kane, and S.~Dawson.
\newblock {\em The Higgs Hunter's Guide}.
\newblock Addison-Wesley, 1990.

\end{thebibliography}
